\documentclass[preprint,review,12pt]{elsarticle}

\usepackage{lineno}   
\usepackage{graphicx}
\usepackage{amsmath,amssymb}
\usepackage{hyperref}
\usepackage{cancel}

\usepackage{color}
\usepackage[dvipsnames]{xcolor}

\usepackage{amssymb}
\usepackage{lipsum}
\usepackage{comment}
\usepackage{ulem}

\journal{Physics Letters B}

\begin{document}

\begin{frontmatter}

\title{Baryon fluctuation signatures
of the onset of deconfinement}

\author[Marek]{Marek Gazdzicki}
\affiliation[Marek]{organization={Jan Kochanowski University},
            city={Kielce},
            country={Poland}}
\author[Mark]{Mark I. Gorenstein}
\affiliation[Mark]{organization={Bogolyubov Institute for Theoretical Physics},
            city={Kyiv},
%            postcode={}, 
%            state={},
            country={Ukraine}}

\author[Anar]{Anar Rustamov}
\affiliation[Anar]{organization={GSI Helmholtzzentrum f{\"u}r Schwerionenforschung},
%            addressline={}, 
            city={Darmstadt},
%            postcode={}, 
%            state={},
            country={Germany}}
\begin{abstract}

An anomalous collision-energy dependence of proton number fluctuations is predicted as a consequence of the onset of deconfinement in heavy-ion collisions at the center of mass energy of the nucleon pair $\sqrt{s_{NN}} \approx 10$ GeV. The effect arises from changes in the effective degrees of freedom between confined and deconfined matter. This may provide a natural explanation of the recent beam-energy-scan results on proton number fluctuations 
at the BNL RHIC and offers a consistent interpretation of data from the CERN SPS and RHIC experiments in terms of the onset of deconfinement.
\end{abstract}

\begin{keyword}
heavy-ion collisions \sep onset of deconfinement \sep fluctuations \sep baryons and protons

\end{keyword}

\end{frontmatter}

\section{Introduction}
\label{sec:intro}

The discovery of the high-density state of strongly interacting matter -- the quark-gluon plasma (QGP) ~\cite{Shuryak:1980tp} -- has been the primary goal of the experimental studies of high-energy nucleus-nucleus collisions. One of the key results was the observation of
the rapid changes in hadron production properties in central Pb+Pb collisions at  
$\sqrt{s_{NN}} \approx 8$~GeV by the NA49 experiment~\cite{NA49:2002pzu, NA49:2007stj}. This collision energy 
was expected~\cite{Gazdzicki:1998vd} for the onset of deconfinement -- the beginning of the creation of a deconfined state of strongly interacting matter with increasing collision energy. These experimental results were confirmed by the RHIC beam energy program~\cite{STAR:2017sal}, and the LHC results support their interpretation (see Ref.~\cite{Rustamov:2012np} and references therein). 

In parallel, the search for the  conjectured critical point -- the endpoint of the first order phase transitions in the QCD phase diagram -- has been  pursued at the CERN SPS and BNL RHIC; see Refs.~\cite{Gazdzicki:2020jte,Bzdak:2019pkr} for reviews.
The collision energy dependence of proton number fluctuations has been measured over a broad range of energies 
by STAR~\cite{STAR:2021iop,STAR:2013gus} at RHIC (BNL), 
HADES~\cite{HADES:2020wpc} at the SIS (GSI), and 
ALICE~\cite{ALICE:2019nbs,ALICE:2022xpf} at the CERN LHC.
The results indicate an anomalous energy dependence in central Au+Au collisions (see Ref.
\cite{Stephanov:2008qz}). 
Hydrodynamics-based calculations 
of the experimental data on proton number fluctuations also suggest possible critical point effects at  $\sqrt{s_{NN}} \le  10$~GeV. More recently, it has been proposed that baryon--baryon interactions undergo a transition from predominantly repulsive to predominantly attractive at around $\sqrt{s_{NN}} \approx 10$~GeV
\cite{Friman:2025swg}. 
In addition, modifications of baryon stopping across the onset of deconfinement were investigated in~\cite{Savchuk:2024ykb}, where they were shown to have a significant impact on baryon number fluctuations.

In the present work, we 
 predict the collision-energy dependence of fluctuations in the net-baryon and proton numbers near the onset of deconfinement. The ideal hadron resonance gas (HRG) and quark-gluon plasma (QGP) approaches are used to model confined and deconfined systems produced in heavy-ion collisions. First, the collision energy dependence of net-baryon number cumulants is calculated within the grand canonical ensemble (GCE). Then, predictions for limited momentum acceptance and global net-baryon conservation are obtained. Finally, we compute predictions for proton-number cumulants.

The paper is organised as follows.
The basic idea is introduced in Sec.~\ref{sec:OoD}. The quantitative model is formulated, and its predictions are presented in Sec.~\ref{sec:fluctuations}. The discussion and conclusions close the paper in Sec.~\ref{sec:conclusions}.

\section{Intuitive arguments}
\label{sec:OoD}
It was suggested many years ago~\cite{Asakawa:2000wh,Jeon:2000wg} that the fluctuations of net baryon number and electric charge may be suppressed when deconfined matter is created.
The idea explores the fact that, in confined matter, the units of baryon number and electric charge are unity, 
whereas in deconfined matter they are carried by quarks with fractional values: $1/3$ for baryon number and $1/3$ or $2/3$ for electric charge.

To illustrate the idea, we first discuss toy models of confined and deconfined matter, considering two systems within the GCE. 
The first system is filled with an ideal Boltzmann gas of baryons, 
while the second contains an  ideal Boltzmann gas of quarks. 
Thus, the quark and baryon number distributions are are Poissonian. Consequently,
all cumulants~\footnote{
In this paper, the first four cumulants are used. They are defined as: 
$\kappa_1[N] = E[N]$ (mean), 
$\kappa_2[N] = E[(N - E[N])^2]$ (variance), 
$\kappa_3[N] = E[(N - E[N])^3]$ (skewness numerator),
$\kappa_4[N] = E[(N - E[N])^4] - 3  E[(N - E[N])^2]^2$ (excess kurtosis numerator). 
The $E[N]$ stands for the expectation value of $N$.
} 
coincide with the mean, and for the baryon gas 
all baryon number cumulant ratios are equal to unity.

Since baryons are formed from triplets of quarks, for sufficiently large number of quarks, the numbers of quarks and baryons are related by
$N_b \approx N_q/3$. Consequently, 
the ratios of baryon number cumulants are 
\begin{align}
\frac{\kappa_2[N_b]}{\kappa_1[N_b]} \approx \frac{1}{3}~,~~~ %\label{eq:DM2} \\[4pt]
\frac{\kappa_3[N_b]}{\kappa_2[N_b]} \approx \frac{1}{3}~,~~~ %\label{eq:DM3} \\[4pt]
\frac{\kappa_4[N_b]}{\kappa_2[N_b]} \approx \frac{1}{9}~. \label{eq:DM-toy}
\end{align}
Therefore, one expects that the energy dependence of cumulant ratios in heavy-ion collisions changes qualitatively during the transition from a confined to a deconfined system.

In addition to the fractional baryon charges of quarks, several important differences arise after deconfinement. First, the relevant degrees of freedom become light quarks whose masses are small compared with the system temperature, whereas the proton mass in the hadronic phase is much larger than the temperature. Second, a substantial population of antiquarks appears, while in the hadronic phase before the onset of deconfinement, the number of antibaryons is negligible. Third, Fermi–Dirac statistics plays an important role for the quark–antiquark system, whereas quantum-statistical effects are negligible for baryons.
All these effects are quantified in the next section.

For completeness, we note that motivated by the ideas from Refs.~~\mbox{\cite{Asakawa:2000wh, Jeon:2000wg}}, net-electric charge fluctuations where measured and discussed in numerous 
%\sout{papers} \corrA{
studies. Here, we refer to the pioneering 
%\sout{studies related to} \corrA{
measurements by NA49~\cite{Zaranek:2001di,NA49:2004fqq} 
%\sout{and the recent ones related to} \corrA{
as well as more recent results from ALICE~\cite{ALICE:2012xnj}.
Significant uncertainties in modelling other processes that influence fluctuations have obscured conclusions regarding the observation of fractional electric charges~\cite{Parra:2025fse}.
It is even more challenging to observe fractional baryonic charges at the CERN LHC energies.

\section{Baryon number fluctuations as the signal of the onset of deconfinement}
\label{sec:fluctuations}
We consider central heavy-ion collisions (Pb+Pb or Au+Au), neglecting, for simplicity, event-by-event fluctuations of the volume of the created system.
We postulate that at the early stage of collisions at low energies ($\sqrt{s_{NN}} \leq 8$~GeV),  the equilibrium HRG 
is created. In contrast,  
for energies $\sqrt{s_{NN}} \geq 12$~GeV~\cite{Gazdzicki:1998vd, Gazdzicki:2014pga}, the equilibrium QGP is formed.

Experiments measure the collision-energy dependence of fluctuations of proton and net-proton numbers.
The latter is defined as the difference between proton and antiproton numbers. 
The results are obtained within a limited momentum acceptance.
To calculate the corresponding predictions for the onset of deconfinement, we proceed in three steps. First, we calculate predictions for net-baryon number fluctuations in the GCE.
Second, we account for the effect of global net-baryon conservation in a limited momentum acceptance.
Third, we estimate the collision-energy dependence of proton-number cumulant ratios.

The GCE predictions for net-baryon fluctuations at 
collision energies $4\le \sqrt{s_{NN}}\le 8$~GeV are calculated using the ideal-gas classical HRG\footnote{At $\sqrt{s_{NN}}\le 4$~GeV, one expects the perceptible contribution to the baryon number fluctuations from nucleon clusters.}. Fermi statistics and excluded-volume repulsion lead to only weak modifications of the results.  
We therefore assume Poisson distributions for the baryon and antibaryon numbers, $N_b$ and $N_{\overline{b}}$. This leads to a Skellam distribution for the net-baryon number, $B = N_b - N_{\overline{b}}$.

\begin{figure}[htb]

\begin{center}
\includegraphics[width=0.6\textwidth]{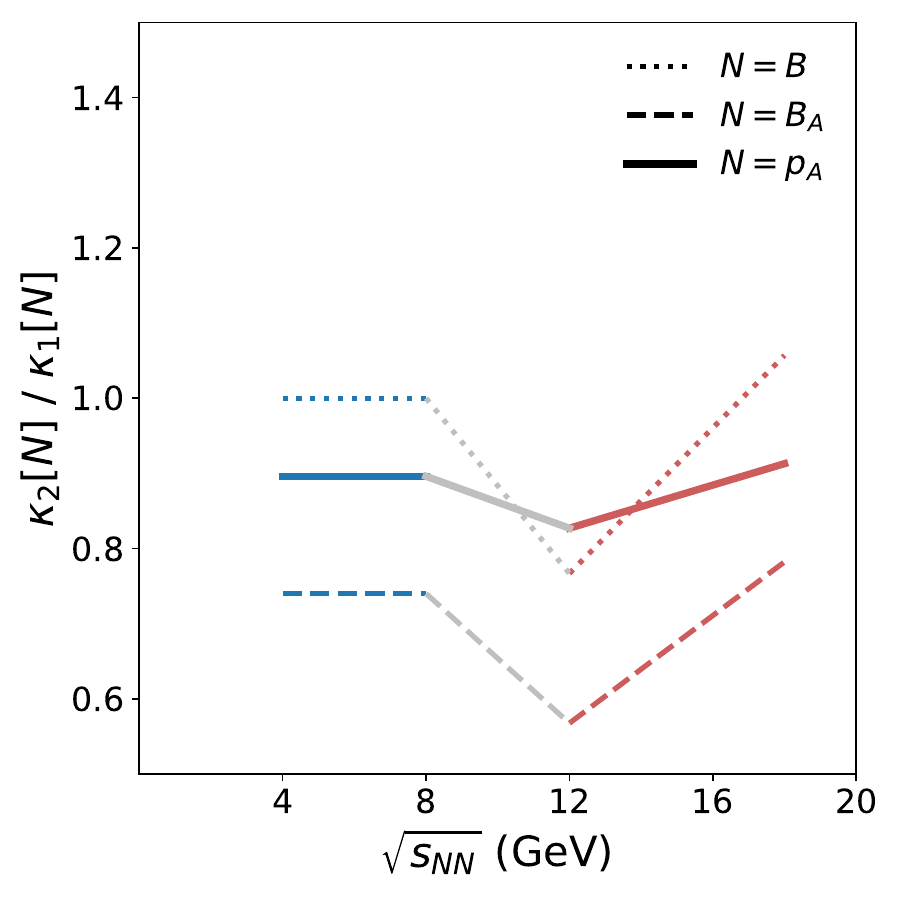} 
\end{center}
\caption{Collision energy dependence of the $\kappa_2[N]/\kappa_1[N]$ ratio predicted in the vicinity of the onset of deconfinement in heavy-ion collisions. Cumulants are calculated for the net baryon number within GCE ($N=B$), the net-baryon number within momentum acceptance taking into account net-baryon number conservation ($N=B_A$), and the proton number within the momentum acceptance ($N=p_A$). 
The lines indicated by light gray colour connect the HRG and QGP dependences across the changeover region to guide the eye.
}
\label{fig:k2k1}
\end{figure}

\begin{figure}[htb]
\begin{center}
\includegraphics[width=0.6\textwidth]{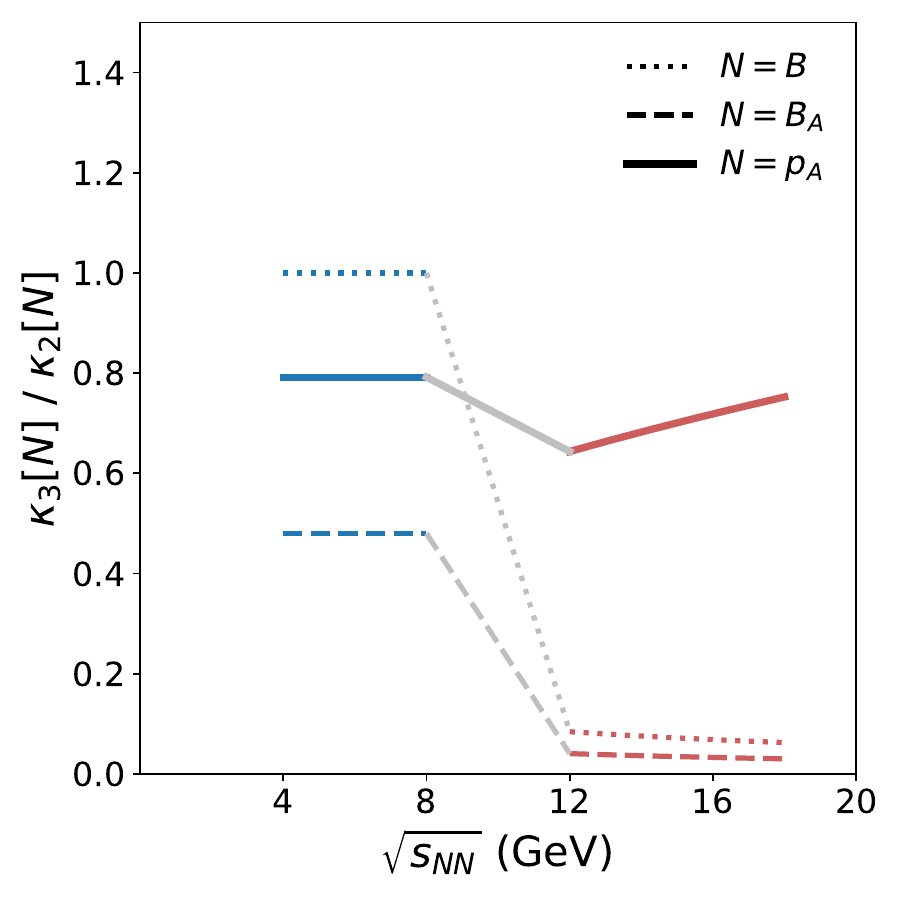} 
\end{center}
\caption{The same as in Fig.~\ref{fig:k2k1} but for $\kappa_3[N]/\kappa_2[N]$
}
\label{fig:k3k2}
\end{figure}

\begin{figure}[htb]
\begin{center}
\includegraphics[width=0.6\textwidth]{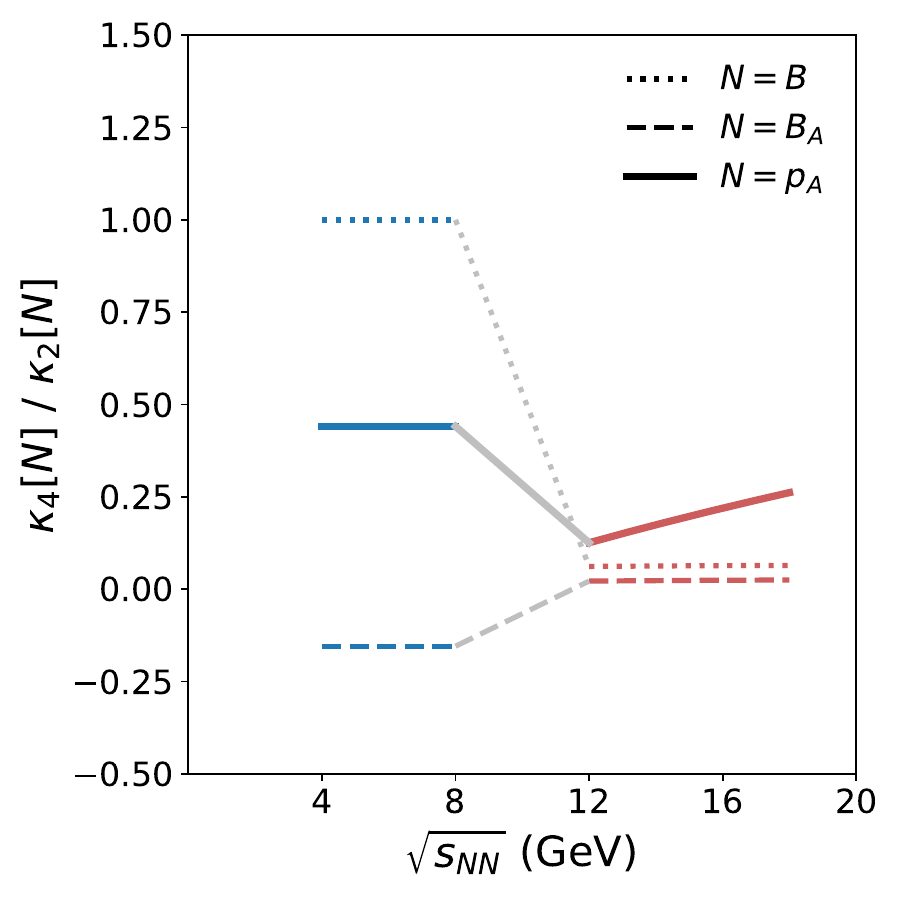}
\end{center}

\caption{The same as in Fig.~\ref{fig:k2k1} but for $\kappa_4[N]/\kappa_2[N]$.
}
\label{fig:k4k2}

\end{figure}

At $\sqrt{s_{NN}} \le 8$~GeV, the antibaryon-to-baryon ratio is below 1$\%$~\cite{Becattini:2005xt}. Therefore, the Skellam distribution of the net-baryon number is close to the Poisson distribution of baryons, which leads to

\begin{align}
\frac{\kappa_2[B]_{H}}{\kappa_1[B]_{H}} = 1~,~~~
\frac{\kappa_3[B]_{H}} {\kappa_2[B]_{H}}= 1~,~~~ 
\frac{\kappa_4[B]_{H}}{\kappa_2[B]_{H}} = 1~, \label{eq:CM4}
\end{align}
for  the net-baryon-number cumulants ratios at $\sqrt{s_{NN}}\le 8$~GeV. 
The subscript $_H$ denotes ratios calculated within the HRG model. The ratios are shown in Figs.~\mbox{\ref{fig:k2k1} - \ref{fig:k4k2}} for collision energies from 4 to 8~GeV.

The net-baryon number fluctuations for the deconfined matter
are calculated using the ideal-gas QGP equation of state with massless $u$, $d$, and $s$ quarks. 
Within this model, the GCE pressure $P$ as a function of temperature $T$ and quark chemical potentials $\mu_f$ reads~\cite{Kapusta:2006pm}
\begin{equation}
 \frac{P}{T^4}~=~ \sum_{f=u,d,s}\left[\frac{7\pi^2}{60}~+~
 \frac{1}{2}\left(\frac{\mu_f}{T}\right)^2~+~\frac{1}{4\pi^2}
 \left( \frac{\mu_f}{T}\right)^4\right]~,   
\end{equation}
where the chemical potentials of $u$, $d$, and $s$  quarks are related to baryon and strangeness chemical potentials $\mu_B$ and $\mu_S$ as
\begin{equation}\label{mu}
 \mu_u=\frac{1}{3}\mu_B~,~~~\mu_d=\frac{1}{3}\mu_B~,~~~
 \mu_s=\frac{1}{3}\mu_B~-~\mu_S~=~0~.
 \end{equation}
The factor 1/3 in Eq.(\ref{mu}) corresponds to the baryon number of quark. 
We take $\mu_S=\mu_B/3$ to ensure zero net-strangeness, and neglect the small values of $\mu_Q$ (for heavy-ion collisions, a rough estimate is $\mu_Q\approx - \mu_B/30$~\cite{Lysenko:2024hqp}).
The cumulants $\kappa_n[B]$ of the net-baryon number distribution follow from
derivatives of the pressure with respect to the baryon chemical potential
\begin{equation}
\kappa_n[B]_Q = VT^3\,
\frac{\partial^n}{\partial(\mu_B/T)^n}
\left(\frac{P}{T^4}\right),
\qquad n=1,2,3,4,..
\label{eq:defkappa}
\end{equation}
This yields
\begin{flalign}
\kappa_1[B]_Q = & {VT^3} ~\left[\frac{2}{9} \frac{\mu_B}{T}~+~\frac{2}{81 \pi^2}  \left(\frac{\mu_B}{T}\right)^3\right]~,       
\label{eq:kappa-1} \\
\kappa_2[B]_Q = &{VT^3}~
\left[\frac{2}{9} +\frac{2}{27}\left(\frac{\mu_B}{\pi T}\right)^2+\frac{1}{9}\right]~, 
\label{eq:kappa-2} \\
\kappa_3[B]_Q = & {VT^3}~\left(\frac{4}{27\pi^2}\frac{\mu_B}{T}\right)~,  %
\label{eq:kappa-3} \\
\kappa_4[B]_Q = & {VT^3}~\left(\frac{4}{27\pi^2}~+~\frac{2}{27\pi^2}\right)~. 
\label{eq:kappa-4} 
\end{flalign}
The subscript $_Q$ denotes cumulants calculated for the QGP.
Note that strange quarks do not contribute to the odd cumulants since $\mu_s=0$. However, they do 
contribute to the even cumulants. The last terms in the right-hand side of Eqs.~(\ref{eq:kappa-2}) 
and~(\ref{eq:kappa-4}) arise from strange–antistrange quark fluctuations at $\mu_s = 0$. 

The corresponding cumulant ratios are
\begin{align}
\frac{\kappa_2[B]_Q}{\kappa_1[B]_Q} =~ &
\frac{3}{2\pi z}~\frac{9+2z^2}{9+z^2}~,~~~ 
\label{eq:DM2} \\
\frac{\kappa_3[B]_Q}{\kappa_2[B]_Q} =~ & \frac{4z}{\pi (9+2z^2)}~,~~~~~
\label{eq:DM3} \\
\frac{\kappa_4[B]_Q}{\kappa_2[B]_Q} =~ & \frac{6}{\pi^2(9+2z^2)}~,
\label{eq:DM4} 
\end{align}
where $z\equiv \mu_B/(\pi T)$.
Note that the ratio $\kappa_2[B]/\kappa_1[B]$ diverges as $\mu_B \to 0$. Thus, this ratio is ill-defined when the net-baryon density vanishes, as in the case of collisions at the LHC.

The collision energy dependence of the net-baryon number cumulants ratios for QGP, calculated according to Eqs.~\mbox{(\ref{eq:DM2}) - (\ref{eq:DM4})}, are shown in Figs.~\mbox{\ref{fig:k2k1} - \ref{fig:k4k2}} for collision energies from 12 to 20~GeV.
We assume that the QGP created at the early stage of the collision hadronizes close to the chemical freeze-out line. The cumulant predictions are then obtained using the energy-dependent parametrization of $T$ and $\mu_B$: 
$T=a_1-a_2\mu_B^2-a_3\mu_B^4~$ and $\mu_B=\frac{b_1}{1+b_2\sqrt{s_{NN}}}$~,
with
$a_1=0.152$~GeV, $a_2=0.026$~GeV$^{-1}$,
$a_3= 0.219$~GeV$^{-3}$ and
$b_1=1.310$~GeV, $b_2=0.278$~GeV$^{-1}$.
The parametrisation was obtained by fitting the mean hadron multiplicities produced in central Pb+Pb and Au+Au collisions; see Ref.~\cite{Cleymans:2005xv}. 

In each collision, the total net-baryon number is conserved.
However, the net-baryon number varies from collision to collision, if it is measured in a limited momentum acceptance.
Within the statistical approach, the effect of net-baryon number conservation in the limited-acceptance region can be modelled using the sub-ensemble acceptance method (SAM) introduced in Ref.~\cite{Vovchenko:2020tsr}.  It relates the GCE predictions to the corresponding expectations for a limited acceptance.
For the considered cumulant ratios, one obtains

\begin{align}
\frac{\kappa_2[B_A]}{\kappa_1[B_A]} = & ~(1-\alpha) ~\frac{\kappa_2[B]}{\kappa_1[B]}~, 
\label{eq:BA2} \\
\frac{\kappa_3[B_A]}{\kappa_2[B_A]}
 = & ~(1-2\alpha) ~\frac{\kappa_3[B]}{\kappa_2[B]}~, 
\label{eq:BA3} \\
\frac{\kappa_4[B_A]}{\kappa_2[B_A]}
 = & ~[1- 3\alpha(1-\alpha)]
~\frac{\kappa_4[B]}{\kappa_2[B]}~ 
 - 3 \alpha ( 1 - \alpha )  ~\left(\frac{\kappa_3[B]}{\kappa_2[B]} \right)^2~,
\label{eq:BA4}
\end{align}
where $B_A$ is the net-baryon number in the momentum acceptance and
$\alpha$ is the fraction of baryons in this acceptance.
The relations were derived considering a limited acceptance in configuration space~\cite{Vovchenko:2020tsr}. Obviously, the SAM method is also valid for particle selection in momentum space for models with uncorrelated particle production in both configuration and momentum spaces.  
In a more general situation, the applicability of the SAM method depends on the presence of a correlation between coordinate and momentum space, for example, due to the collective flow of matter; see 
Ref.~\cite{Kuznietsov:2024xyn} for details.  
In the absence of an alternative method, we employ the SAM method to estimate the predictions.

Figures~\mbox{\ref{fig:k2k1} - \ref{fig:k4k2}}
show predictions for the $B_A$ cumulant ratios obtained using Eqs.~\mbox{(\ref{eq:BA2}) - (\ref{eq:BA4})}. These equations are applied to the previously calculated predictions for $B$ cumulants obtained within GCE, i.e. Eq.~(\ref{eq:CM4}) for the HRG  and Eqs.~\mbox{(\ref{eq:DM2}) - (\ref{eq:DM4})} for the QGP.
Here, a typical experimental value of the parameter $\alpha = 0.26$ was used~\cite{Friman:2025swg}.

The experiments measure fluctuations of proton number rather than of baryon number. To relate previously calculated net-baryon-number cumulant ratios in the acceptance to proton-number ratios, a hadronization model is required.
At the considered collision energies ($\sqrt{s_{NN}} \lesssim 20$~GeV) the mean number of anti-baryons is smaller than 10\% of the mean number of baryons~\cite{Becattini:2005xt}. Thus, we approximate the net-baryon-number cumulants by the baryon-number cumulants. Furthermore, we 
assume that for a fixed net-baryon number $B_A$, the proton number $p_A$ fluctuates according to a binomial distribution with the probability $\beta$ -- the so-called binomial thinning. Under the binomial thinning, the cumulants are modified according to the standard relations
given in~\ref{app:thin}.

Figures~\ref{fig:k2k1} - \ref{fig:k4k2} show predictions for the proton number cumulants calculated for the HRG and the QGP using Eqs.~(\ref{eq:p2}) - (\ref{eq:p4}), applied to the previously calculated prediction for $B_A$ cumulants, Eqs.~(\ref{eq:BA2}) - (\ref{eq:BA4}).
Here, a typical value of the parameter $\beta = 0.4$ was used~\cite{Becattini:2005xt}.

\section{Discussion and conclusions}
\label{sec:conclusions}

The collision energy dependence of the net-baryon number in the GCE ($B$), net-baryon number in the momentum acceptance ($B_A$) and proton number in momentum acceptance ($p_A$) is predicted to change rapidly in the vicinity of the onset of deconfinement ($\sqrt{s_{NN}} \approx 8 - 12$~GeV). This is demonstrated in Figs.~\mbox{\ref{fig:k2k1} - \ref{fig:k4k2}}. The largest changes are observed for $B$ and the smallest for $p_A$. The magnitude of the change in the proton number cumulants ratios significantly depends on the fraction of accepted baryons ($\alpha$) and the fraction of protons among the accepted baryons ($\beta$). 
In Appendix~\ref{app:data}, we present the collision-energy dependence of the proton-number factorial cumulant ratios measured in the STAR experiment together with the predictions obtained in this paper for the onset of deconfinement.
The model calculations are done for parameters $\alpha = 0.26$ and $\beta = 0.4$, the typical experimental values for STAR~\cite{Braun-Munzinger:2020jbk, STAR:2021iop}. Both the data and the model exhibit a qualitative change in the collision-energy dependence at $\sqrt{s_{NN}} \approx 8$--12~GeV.

A quantitive comparison of the onset of deconfinement predictions with the experimental results requires, in particular,
\begin{enumerate}[(i)]
    \item 
    using experimentally estimated values of the parameter $\alpha$, which significantly varies between data points~\cite{Braun-Munzinger:2020jbk},
    \item 
    using experimentally estimated values of the parameter $\beta$, which varies between data points,
    \item 
    improved modelling of fluctuations of the fraction of protons for a given number of accepted baryons,
    \item 
    taking into account the production of nuclear clusters, 
    \item 
    taking into account the nuclear critical point which can significantly enhance the baryon number fluctuations at low collision energies~\cite{Poberezhnyuk:2019pxs}.
    \item 
    studying the sensitivity of the predictions to the QGP equation of state,
    \item 
    investigating the sensitivity of the results to the assumption that net-baryon fluctuations freeze out in the vicinity of the chemical freeze-out line.
\end{enumerate}

The results presented here open the possibility for a consistent interpretation of experimental data from GSI SIS through CERN SPS to BNL RHIC and the CERN LHC.

\section*{Acknowledgements}
The authors thank Stanislaw Mrowczynski, Volker Koch, and Volodymyr Vovchenko for reading the manuscript and providing fruitful comments..
This work was supported by the Polish National Science Centre (NCN) grant 2018/30/A/ST2/00226 
and the National Research Foundation of Ukraine under grant 2025.07/0050.  

\appendix

\section{The binomial thinning}
\label{app:thin}

It is assumed that for a fixed number of baryons, the proton number fluctuates according to a binomial distribution with the probability $\beta$ -- the so-called binomial thinning. Under the binomial thinning, the cumulants are modified according to the relations:
\begin{align}
   \frac{\kappa_2[p_A]}{\kappa_1[p_A]} &= 1 - \beta + \beta ~
   \frac{\kappa_2[B_A]}{\kappa_1[B_A]}~,    
\label{eq:p2} \\[4pt]
\frac{\kappa_3[p_A]}{\kappa_2[p_A]} &= \frac{\kappa_2[B_A] /\kappa_1[B_A]} {\kappa_2[p_A] /\kappa_1[p_A]} 
\left[ \beta^2 \frac{\kappa_3[B_A]}{\kappa_2[B_A]} + 3\beta (1-\beta)\right] 
+ \frac{1-\beta}{\kappa_2[p_A]/\kappa_1[p_A]} (1-2\beta)~,
\label{eq:p3} \\[4pt]
\frac{\kappa_4[p_A]}{\kappa_2[p_A]}
&=
\frac{\kappa_2[B_A]/\kappa_1[B_A]}{\kappa_2[p_A]/\kappa_1[p_A]}
\Biggl[
\beta^3 \frac{\kappa_4[B_A]}{\kappa_2[B_A]}
\nonumber\\
&\qquad
+ 6 \beta^2 (1-\beta)
\frac{\kappa_3[B_A]}{\kappa_2[B_A]}
+ \beta(1-\beta)(7-11\beta)
\Biggr]
\nonumber\\
&\qquad
+ \frac{1-\beta}{\kappa_2[p_A]/\kappa_1[p_A]}
\bigl[1-6\beta(1-\beta)\bigr]~.
\label{eq:p4} 
\end{align}

\section{Comparison with experimental results}
\label{app:data}

\begin{figure}[!t]

\begin{center}
\includegraphics[width=0.4\textwidth]{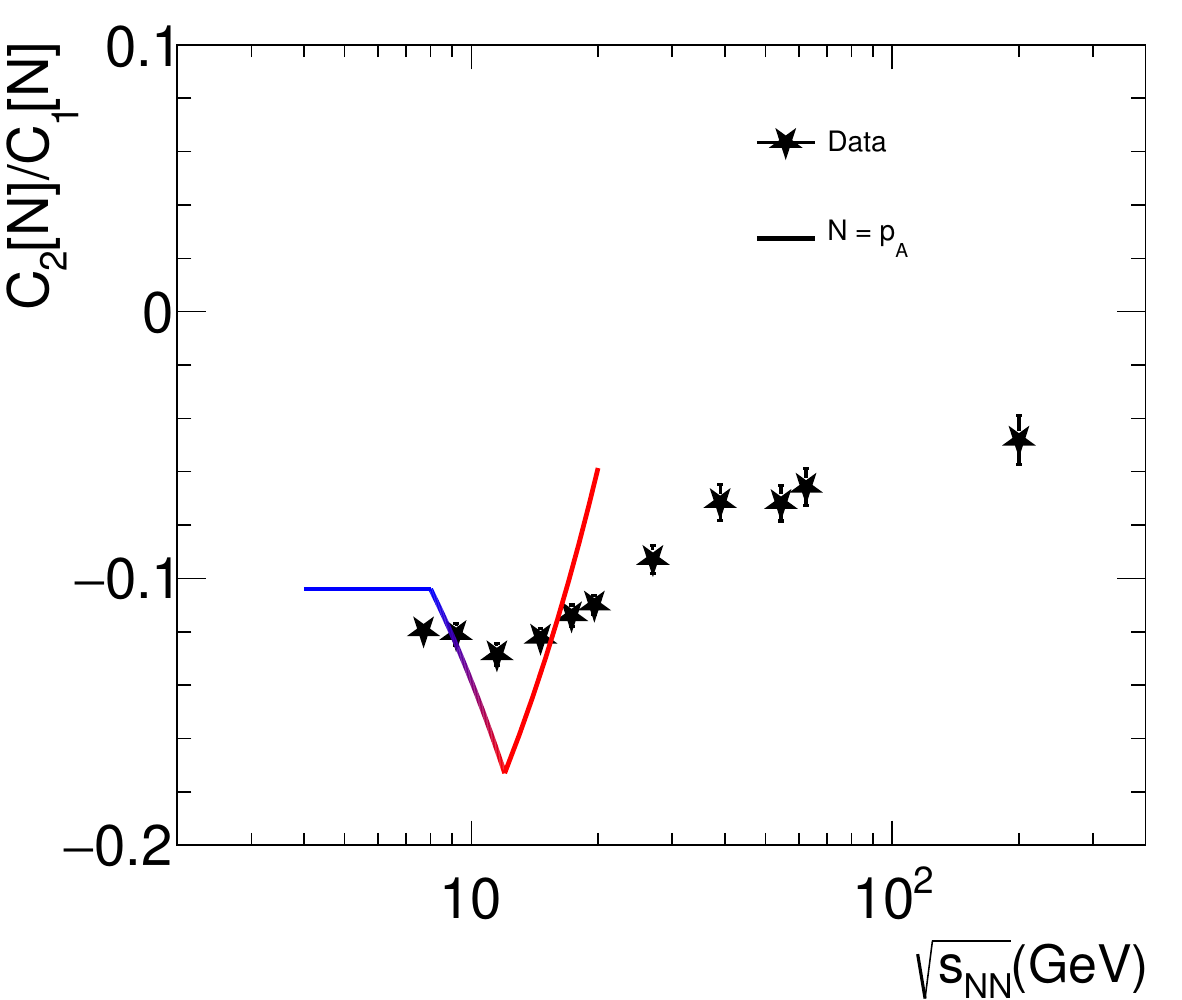} 
\includegraphics[width=0.4\textwidth]{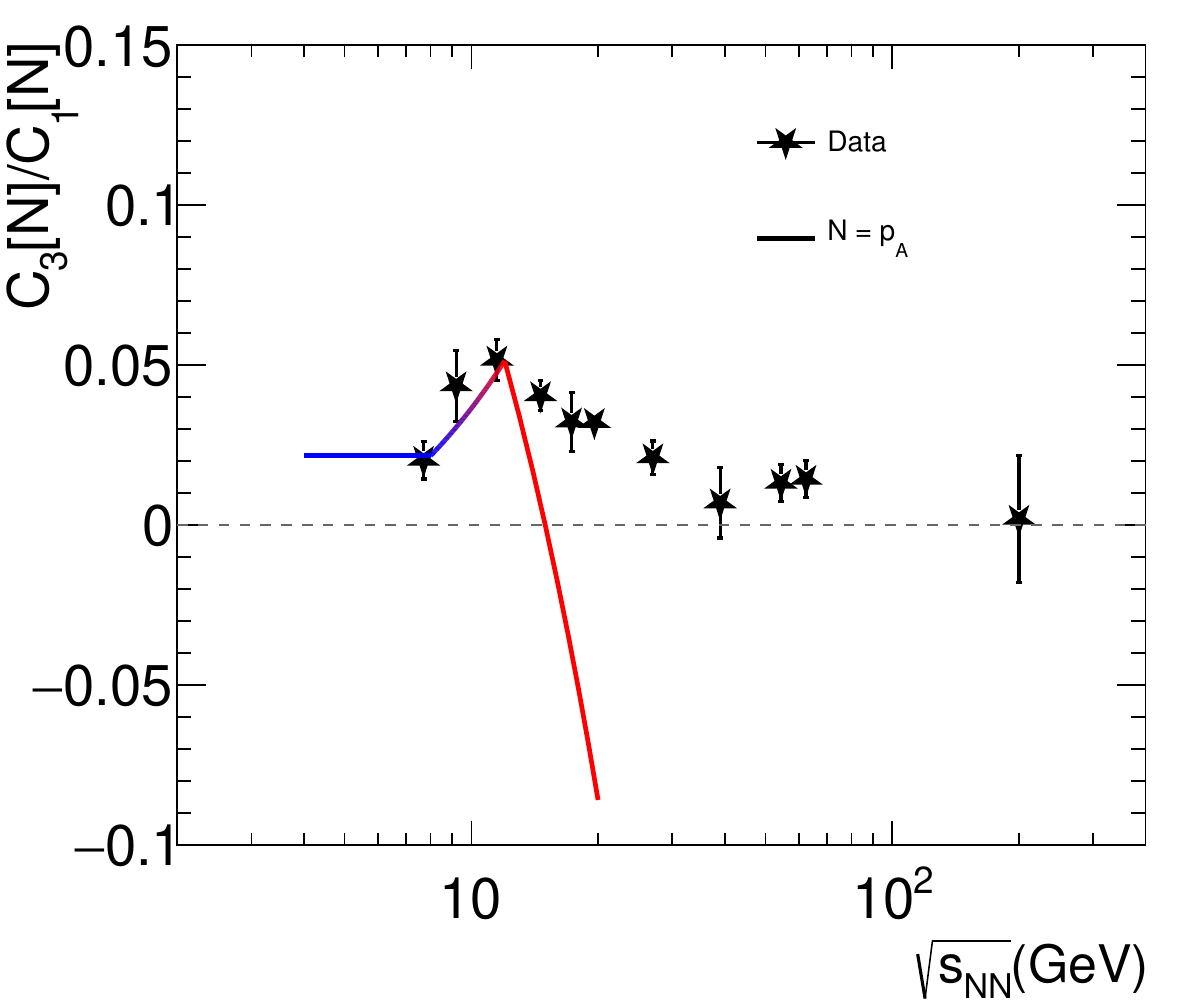} 
\includegraphics[width=0.4\textwidth]{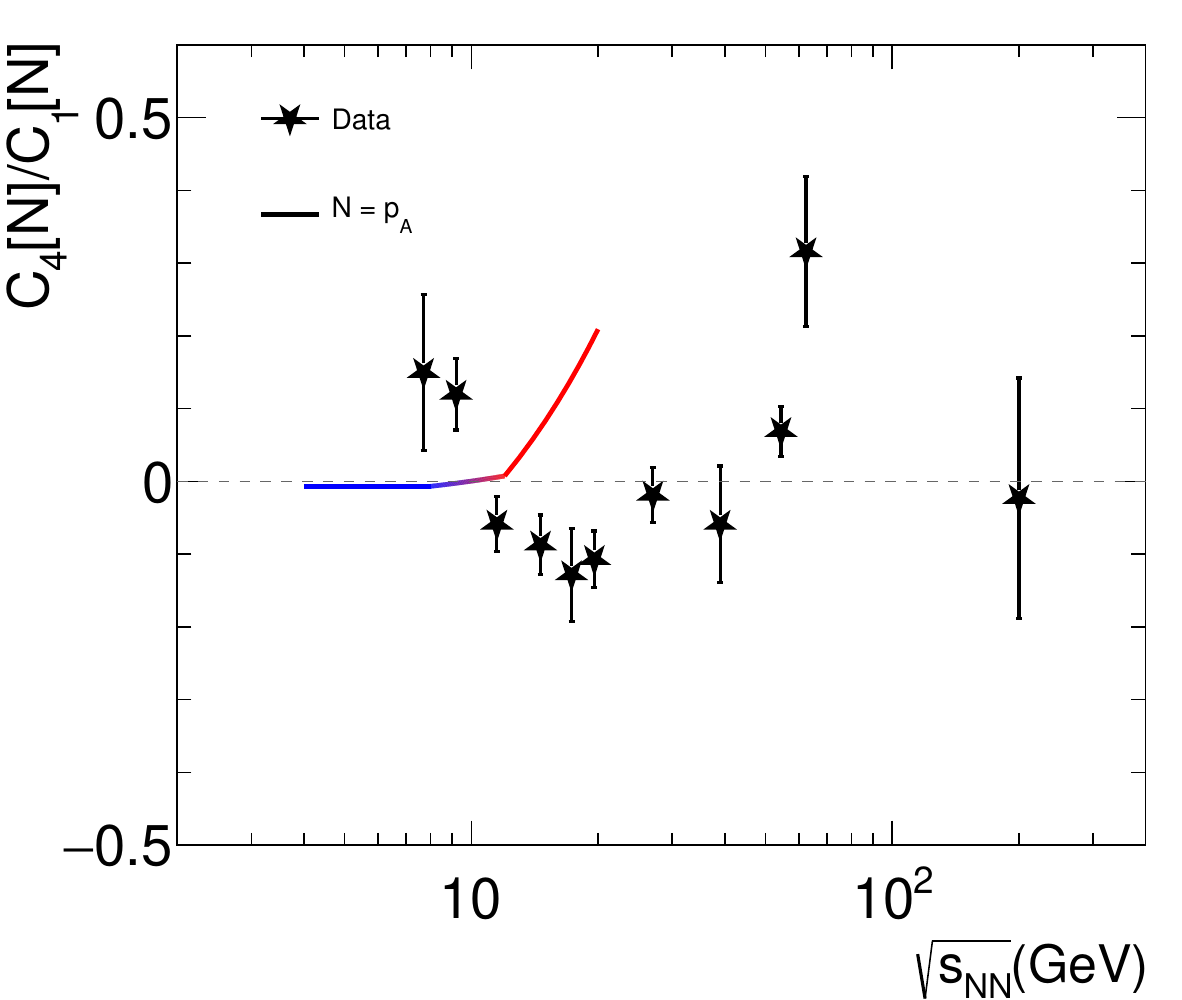} 
\end{center}
\caption{Collision-energy dependence of the factorial cumulant ratios of the proton number in the acceptance. Symbols show measurements from the STAR experiment~\cite{STAR:2025zdq, Braun-Munzinger:2026krf}, while solid lines correspond to our model calculations. 
}
\label{fig:fact_cums}
\end{figure}

In this section, we compare the predictions of our model with the measured factorial cumulant ratios of the proton number reported by the STAR collaboration~\cite{STAR:2021iop}.

The relation between factorial cumulants $C_{n}$ and ordinary cumulants $\kappa_{n}$ 
is mediated by the signed Stirling numbers of the first kind $s(n,j)$

\begin{equation}
    C_{n} = \sum_{j=1}^{n} s(n,j)\,\kappa_{j},
    \label{eq_stirling2}
\end{equation}
For the first four cumulants Eq.~\ref{eq_stirling2} yields
\begin{eqnarray}
    C_{1} &=& \kappa_{1}\\
    C_{2} &=& -\kappa_{1} + \kappa_{2},\\
    C_{3} &=& 2\kappa_{1} - 3\kappa_{2} + \kappa_{3},\nonumber\\
    C_{4} &=& -6\kappa_{1} + 11\kappa_{2} - 6\kappa_{3} + \kappa_{4}\nonumber.
\end{eqnarray}

The considered cumulants ratios are

\begin{eqnarray}
    C_{2}/C_{1} &=& -1 + \kappa_{2}/\kappa_{1}, \nonumber\\
    C_{3}/C_{1} &=& 2 -3\kappa_{2}/\kappa_{1} + \kappa_{3}/\kappa_{1},\\
    C_{4}/C_{1} &=& -6 + 11\kappa_{2}/\kappa_{1} - 6\kappa_{3}/\kappa_{1} + \kappa_{4}/\kappa_{1}\nonumber.
\end{eqnarray}

Fig.~\ref{fig:fact_cums} shows the collision-energy dependence of the proton-number factorial cumulants measured by the STAR experiment~\cite{STAR:2025zdq,Braun-Munzinger:2026krf} (symbols) in comparison with our model calculations (solid lines). More specifically, our calculations for the equilibrium HRG are shown by blue lines for energies between 4 and 8~GeV. For energies above 12~GeV, the calculations correspond to the equilibrium QGP case. In general, the non-monotonic behaviour observed in the energy dependence of the measured cumulant ratios is qualitatively captured by the model predictions. For a more quantitative comparison, however, the remarks made in Section~\ref{sec:conclusions} need to be considered; these will be addressed in our future studies.

\bibliographystyle{unsrt}
\bibliography{main_ref.bib}

\end{document}